\documentclass[sigconf]{acmart}
\usepackage{booktabs} 
\usepackage{multirow}
\usepackage{color}
\usepackage{subfigure} 
\usepackage{array}
\usepackage{breqn}
\usepackage{{inputenc}}
\usepackage{balance}
\usepackage{multirow,tabularx}
\usepackage{longtable}
\usepackage{booktabs,longtable}
\usepackage{hhline}
\usepackage[flushleft]{threeparttable}
\usepackage{algorithm}
\usepackage{algorithmic}
\usepackage{epsfig}
\usepackage{graphicx}
\usepackage{epstopdf}
\usepackage{thmtools}
\usepackage{enumitem}
\usepackage{verbatim}
\usepackage{amsfonts}
\usepackage{hyperref}
\hypersetup{colorlinks=false,linkcolor=blue,urlcolor=blue,citecolor=red}
\newcommand\norm[1]{\left\lVert#1\right\rVert}
\newcommand{\Lapl}{\mathbf{\mathop{\mathcal{L}}}}
\newcommand{\Trans}[1]{{#1}^{\top}}

\newcommand{\Mat}[1]{\mathbf{#1}}

\newcommand{\Space}[1]{\mathbb{#1}}
\newcommand{\Set}[1]{\mathcal{#1}}

\newcommand{\BlockMatSquare}[4]{\left[\begin{matrix}#1 & #2\\#3 & #4\end{matrix}\right]}

\newcommand{\ie}{\emph{i.e., }}
\newcommand{\eg}{\emph{e.g., }}

\newcommand{\etc}{\emph{etc.}}
\newcommand{\wrt}{\emph{w.r.t. }}
\newcommand{\cf}{\emph{cf. }}
\newcommand{\aka}{\emph{aka. }}

\theoremstyle{definition}

\def\BibTeX{{\rm B\kern-.05em{\sc i\kern-.025em b}\kern-.08emT\kern-.1667em\lower.7ex\hbox{E}\kern-.125emX}}

\begin{document}
\settopmatter{printacmref=true}
\fancyhead{}

\title{Neural Graph Collaborative Filtering}
\titlenote{In the version published in ACM Digital Library, we find some small bugs; the bugs do not change the comparison results and the empirical findings. In this latest version, we update and correct the experimental results (\ie the preprocessing of Yelp2018 dataset and the ndcg metric). All updates are highlighted in footnotes.}

\author{Xiang Wang}
\affiliation{%
	\institution{National University of Singapore}
}
\email{xiangwang@u.nus.edu}

\author{Xiangnan He}
\authornote{Xiangnan He is the corresponding author.}
\affiliation{%
	\institution{University of Science and Technology of China}
}
\email{xiangnanhe@gmail.com}

\author{Meng Wang}
\affiliation{%
	\institution{Hefei University of Technology}
}
\email{eric.mengwang@gmail.com}

\author{Fuli Feng}
\affiliation{%
	\institution{National University of Singapore}
}
\email{fulifeng93@gmail.com}

\author{Tat-Seng Chua}
\affiliation{%
	\institution{National University of Singapore}
}
\email{dcscts@nus.edu.sg}

\begin{abstract}
Learning vector representations (\aka embeddings) of users and items lies at the core of modern recommender systems. Ranging from early matrix factorization to recently emerged deep learning based methods, existing efforts typically obtain a user's (or an item's) embedding by mapping from pre-existing features that describe the user (or the item), such as ID and attributes. We argue that an inherent drawback of such methods is that, the \textbf{collaborative signal}, which is latent in user-item interactions, is not encoded in the embedding process. As such, the resultant embeddings may not be sufficient to capture the collaborative filtering effect.  

In this work, we propose to integrate the user-item interactions --- more specifically the bipartite graph structure --- into the embedding process. We develop a new recommendation framework \textit{Neural Graph Collaborative Filtering} (NGCF), which exploits the user-item graph structure by propagating embeddings on it. 
This leads to the expressive modeling of \textbf{high-order connectivity} in user-item graph, effectively injecting the collaborative signal into the embedding process in an explicit manner. 
We conduct extensive experiments on three public benchmarks,
demonstrating significant improvements over several state-of-the-art models like HOP-Rec~\cite{HOP-rec} and Collaborative Memory Network~\cite{CMN}. 
Further analysis verifies the importance of embedding propagation for learning better user and item representations, justifying the rationality and effectiveness of NGCF.
Codes are available at \url{https://github.com/xiangwang1223/neural_graph_collaborative_filtering}.

\end{abstract}

\copyrightyear{2019}
\acmYear{2019}
\setcopyright{acmcopyright}
\acmConference[SIGIR '19]{Proceedings of the 42nd International ACM SIGIR
Conference on Research and Development in Information Retrieval}{July 21--25,
2019}{Paris, France}
\acmBooktitle{Proceedings of the 42nd International ACM SIGIR Conference on
Research and Development in Information Retrieval (SIGIR '19), July 21--25, 2019,
Paris, France}
\acmPrice{15.00}
\acmDOI{10.1145/3331184.3331267}
\acmISBN{978-1-4503-6172-9/19/07}

%
%

\vspace{-5pt}
\maketitle

\section{Introduction}
Personalized recommendation is ubiquitous, having been applied to many online services such as E-commerce, advertising, and social media.
At its core is estimating how likely a user will adopt an item based on the historical interactions like purchases and clicks.
Collaborative filtering (CF) addresses it by assuming that behaviorally similar users would exhibit similar preference on items. 
To implement the assumption, a common paradigm is to parameterize users and items for reconstructing historical interactions, and predict user preference based on the parameters~\cite{NCF,KTUP}.

Generally speaking, there are two key components in learnable CF models --- 1) \emph{embedding}, which transforms users and items to vectorized representations, and 2) \emph{interaction modeling}, which reconstructs historical interactions based on the embeddings.
For example, matrix factorization (MF) directly embeds user/item ID as an vector and models user-item interaction with inner product~\cite{MF}; collaborative deep learning extends the MF embedding function by integrating the deep representations learned from rich side information of items~\cite{CDL}; 
neural collaborative filtering models replace the MF interaction function of inner product with nonlinear neural networks~\cite{NCF}; and translation-based CF models instead use Euclidean distance metric as the
interaction function~\cite{tay2018latent}, among others.

Despite their effectiveness, we argue that these methods are not sufficient to yield 
satisfactory embeddings for CF. The key reason is that the embedding function lacks an explicit encoding of the crucial \textbf{collaborative signal}, which is latent in user-item interactions to reveal the behavioral similarity between users (or items). To be more specific, most existing methods build the embedding function with the descriptive features only (\eg ID and attributes), 
without considering the
\textit{user-item interactions} --- which are only used to define the objective function for model training~\cite{BPRMF,tay2018latent}. 
As a result, when the embeddings are insufficient in capturing CF, the methods have to rely on the interaction function to make up for the deficiency of suboptimal embeddings~\cite{NCF}. 

\begin{figure}[t]
    \centering
	\includegraphics[width=0.43\textwidth]{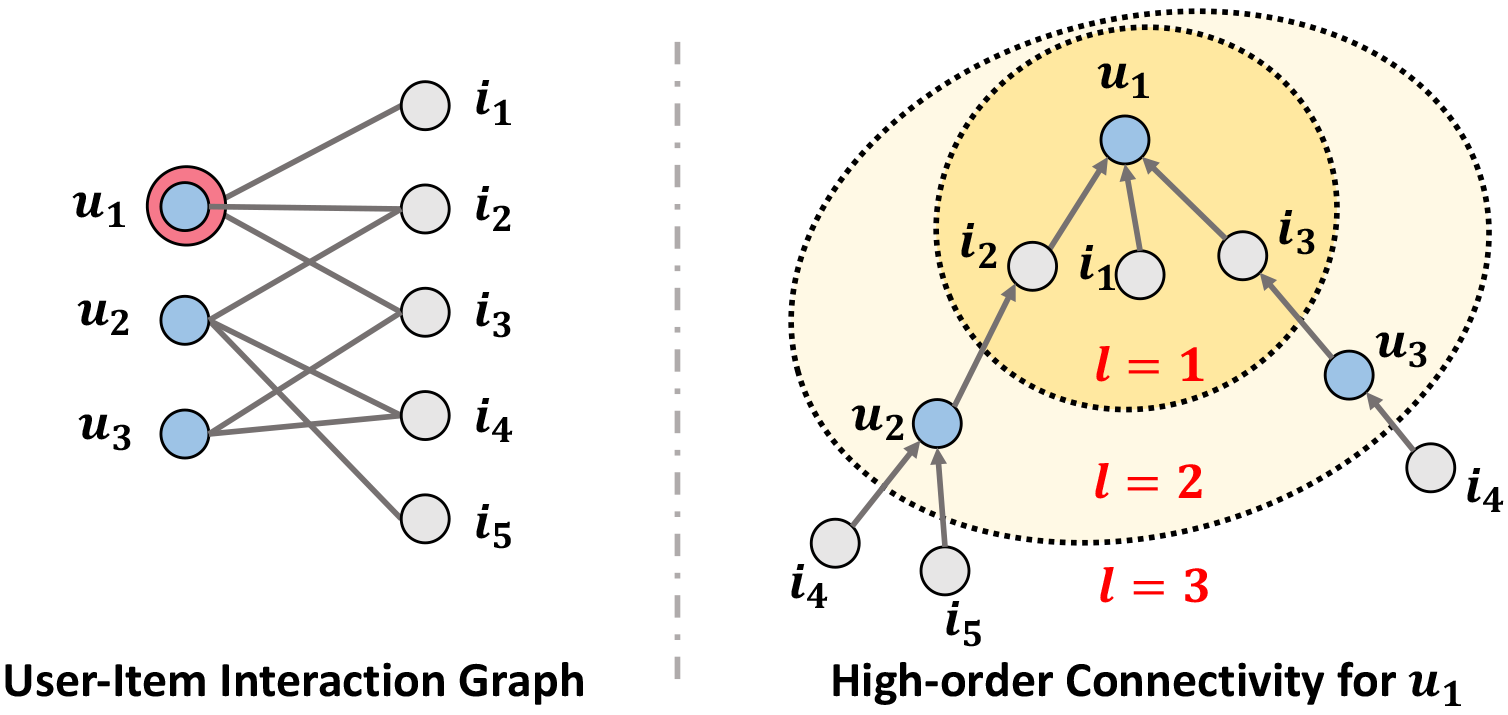}
	\vspace{-10pt}
	\caption{An illustration of the user-item interaction graph and the high-order connectivity. The node $u_1$ is the target user to provide recommendations for.}
	\label{fig:intro}
	\vspace{-15pt}
\end{figure}

While intuitively useful to integrate user-item interactions into the embedding function, it is non-trivial to do it well. In particular, the scale of interactions can easily reach millions or even larger in real applications, making it difficult to distill the desired collaborative signal. 
In this work, we tackle the challenge by exploiting the \textbf{high-order connectivity} from user-item interactions, a natural way that encodes collaborative signal in the interaction graph structure.

\vspace{+5pt}
\noindent\textbf{Running Example}. Figure \ref{fig:intro} illustrates the concept of high-order connectivity. The user of interest for recommendation is $u_1$, labeled with the double circle in the left subfigure of user-item interaction graph. The right subfigure shows the tree structure that is expanded from $u_1$. The high-order connectivity denotes the path that reaches $u_1$ from any node with the path length $l$ larger than 1. Such high-order connectivity contains rich semantics that carry collaborative signal. 
For example, the path $u_{1}\leftarrow i_{2}\leftarrow u_{2}$ indicates the behavior similarity between $u_1$ and $u_2$, as both users have interacted with $i_2$; the longer path $u_{1}\leftarrow i_{2}\leftarrow u_{2}\leftarrow i_4$ suggests that $u_1$ is likely to adopt $i_4$, since her similar user $u_2$ has consumed $i_4$ before. Moreover, from the holistic view of $l=3$, item $i_4$ is more likely to be of interest to $u_1$ than item $i_5$, since there are two paths connecting <$i_4, u_1$>, while only one path connects <$i_5, u_1$>.

\vspace{5pt}
\noindent\textbf{Present Work}. We propose to model the high-order connectivity information in the embedding function. Instead of expanding the interaction graph as a tree which is complex to implement, we design a neural network method to propagate embeddings recursively on the graph. This is inspired by the recent developments of graph neural networks~\cite{GraphSAGE,JumpKG,KGAT},
which can be seen as constructing information flows in the embedding space. Specifically, we devise an \textbf{embedding propagation} layer, which refines a user's (or an item's) embedding by aggregating the embeddings of the interacted items (or users). By stacking multiple embedding propagation layers, we can enforce the embeddings to capture the collaborative signal in high-order connectivities. Taking Figure~\ref{fig:intro} as an example, stacking two layers captures the behavior similarity of $u_1\leftarrow i_2\leftarrow u_2$, stacking three layers captures the potential recommendations of $u_1\leftarrow i_2\leftarrow u_2\leftarrow i_4$, and the strength of the information flow (which is estimated by the trainable weights between layers) determines the recommendation priority of $i_4$ and $i_5$. 
We conduct extensive experiments on three public benchmarks to verify the rationality and effectiveness of our \textit{Neural Graph Collaborative Filtering} (NGCF) method. 

Lastly, it is worth mentioning that although the high-order connectivity information has been considered in a very recent method named HOP-Rec~\cite{HOP-rec}, it is only exploited to enrich the training data. Specifically, the prediction model of HOP-Rec remains to be MF, while it is trained by optimizing a loss that is augmented with high-order connectivities. Distinct from HOP-Rec, we contribute a new technique to integrate high-order connectivities into the prediction model, which empirically yields better embeddings than HOP-Rec for CF. 

To summarize, this work makes the following main contributions:
\begin{itemize}[leftmargin=*]
    \item We highlight the critical importance of explicitly exploiting the collaborative signal in the embedding function of model-based CF methods. 
    
    \item We propose NGCF, a new recommendation framework based on graph neural network, which explicitly encodes the collaborative signal in the form of high-order connectivities by performing embedding propagation.
    
    \item We conduct empirical studies on three million-size datasets. Extensive results demonstrate the state-of-the-art performance of NGCF and its effectiveness in improving the embedding quality with neural embedding propagation.
\end{itemize}
\section{Methodology}

\begin{figure}[t]
\centering
\includegraphics[width=0.42\textwidth]{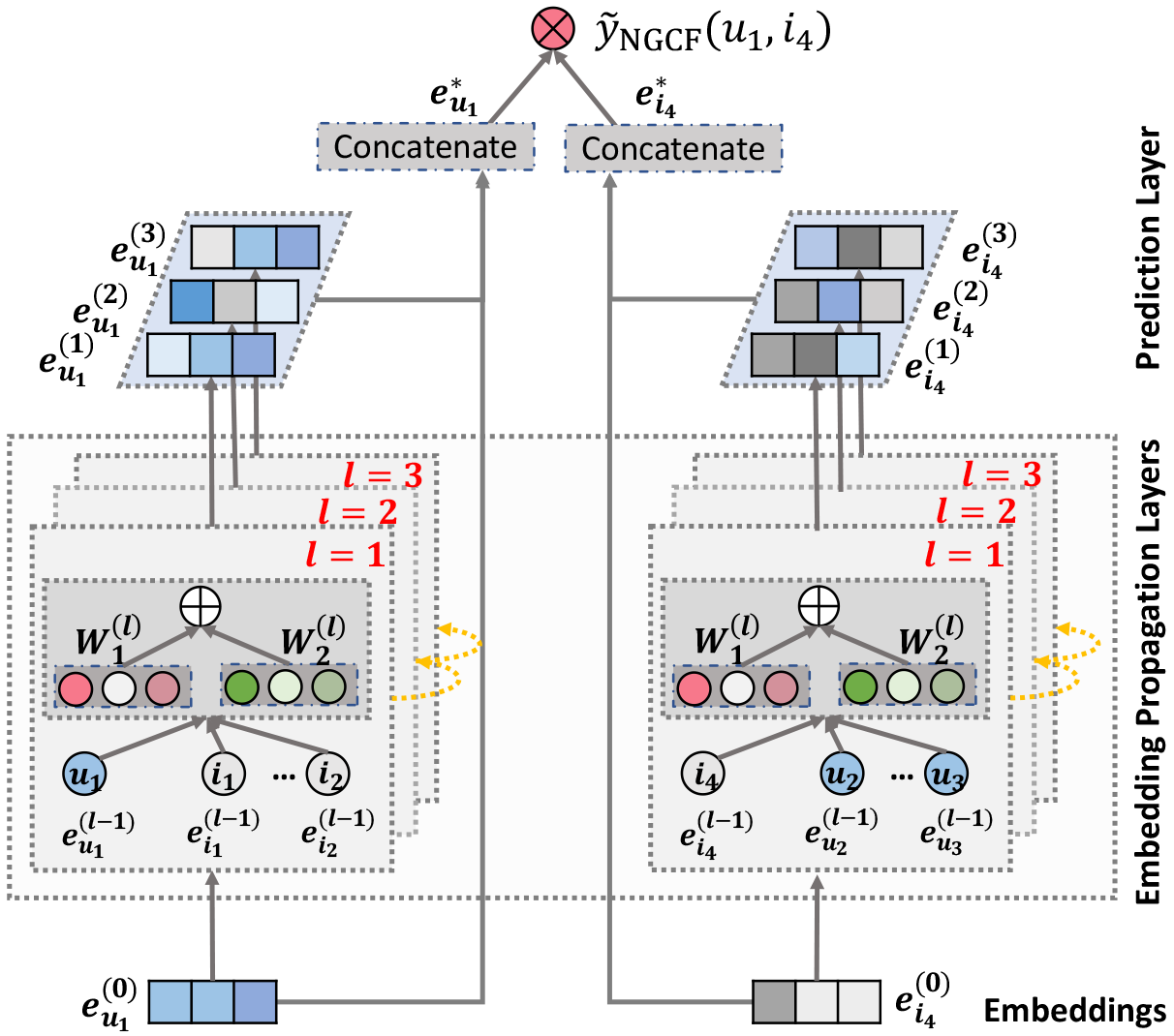}
\vspace{-10pt}
\caption{An illustration of NGCF model architecture  (the arrowed lines present the flow of information). The representations of user $u_{1}$ (left) and item $i_{4}$ (right) are refined with multiple embedding propagation layers, whose outputs are  concatenated to make the final prediction.
}
\label{fig:embedding-propagation}
\vspace{-15pt}
\end{figure}

We now present the proposed NGCF model, the architecture of which is illustrated in Figure~\ref{fig:embedding-propagation}.
There are three components in the framework:
(1) an embedding layer that offers and initialization of user embeddings and item embeddings; 
(2) multiple embedding propagation layers that refine the embeddings by
injecting high-order connectivity relations;
and (3) the prediction layer that aggregates the refined embeddings from different propagation layers and outputs the affinity score of a user-item pair.
Finally, we discuss the time complexity of NGCF and the connections with existing methods.

\subsection{Embedding Layer}
Following mainstream recommender models~\cite{BPRMF,NCF,KTUP}, we describe a user $u$ (an item $i$) with an embedding vector $\Mat{e}_{u}\in\Space{R}^{d}$ ($\Mat{e}_{i}\in\Space{R}^{d}$), where  $d$ denotes the embedding size.
This can be seen as building a parameter matrix as an embedding look-up table: 
\begin{gather}\label{equ:e-0}
	\Mat{E}=[\underbrace{\Mat{e}_{u_{1}},\cdots,\Mat{e}_{u_{N}}}_{\text{users embeddings}}, \underbrace{\Mat{e}_{i_{1}},\cdots,\Mat{e}_{i_{M}}}_{\text{item embeddings}}].
\end{gather}
It is worth noting that this embedding table serves as an initial state for user embeddings and item embeddings, to be optimized in an end-to-end fashion. 
In traditional recommender models like MF and neural collaborative filtering~\cite{NCF}, these ID embeddings are directly fed into an interaction layer (or operator) to achieve the prediction score. 
In contrast, in our NGCF framework, we refine the embeddings by propagating them on the user-item interaction graph. This leads to more effective embeddings for recommendation, since the embedding refinement step explicitly injects collaborative signal into embeddings. 

\subsection{Embedding Propagation Layers}
Next we build upon the message-passing architecture of GNNs~\cite{GraphSAGE,JumpKG} in order to capture CF signal along the graph structure and refine the embeddings of users and items. We first illustrate the design of one-layer propagation, and then generalize it to multiple successive layers. 

\subsubsection{\textbf{First-order Propagation}}
Intuitively, the interacted items provide direct evidence on a user's preference~\cite{DeepICF,FISM};
analogously, the users that consume an item can be treated as the item's features and used to measure the collaborative similarity of two items.
We build upon this basis to perform embedding propagation between the connected users and items, formulating the process with two major operations: \emph{message construction} and \emph{message aggregation}. 

\vspace{+5pt}
\noindent{\textbf{Message Construction}}.
For a connected user-item pair $(u,i)$, we define the message from $i$ to $u$ as:
\begin{gather}
	\Mat{m}_{u\leftarrow i}=f(\Mat{e}_{i},\Mat{e}_{u},p_{ui}),
\end{gather}
where $\Mat{m}_{u\leftarrow i}$ is the message embedding (\ie the information to be propagated). $f(\cdot)$ is the message encoding function, which takes embeddings $\Mat{e}_{i}$ and $\Mat{e}_{u}$ as input, and uses the coefficient $p_{ui}$ to control the decay factor on each propagation on edge $(u,i)$.

In this work, we implement $f(\cdot)$ as:
\begin{gather}\label{equ:0-message}
	\Mat{m}_{u\leftarrow i}=\frac{1}{\sqrt{|\Set{N}_{u}||\Set{N}_{i}|}}\Big(\Mat{W}_{1}\Mat{e}_{i} + \Mat{W}_{2}(\Mat{e}_{i}\odot\Mat{e}_{u})\Big),
\end{gather}
where $\Mat{W}_1,\Mat{W}_2\in\Space{R}^{d'\times d}$ are the trainable weight matrices to distill useful information for propagation, and $d'$ is the transformation size.
Distinct from conventional graph convolution networks~\cite{FirstGCN,GCN,PinSage,GC-MC} that consider the contribution of $\Mat{e}_i$ only, here we additionally encode the interaction between $\Mat{e}_i$ and $\Mat{e}_u$ into the message being passed via $\Mat{e}_i\odot\Mat{e}_u$, where $\odot$ denotes the element-wise product.
This makes the message dependent on the affinity between $\Mat{e}_{i}$ and $\Mat{e}_{u}$, \eg passing more messages from the similar items. This not only increases the model representation ability,
but also boosts the performance for recommendation (evidences in our experiments Section~\ref{sec:layer-effect}). 

Following the graph convolutional network~\cite{GCN}, we set $p_{ui}$ as the graph Laplacian norm $1/\sqrt{|\Set{N}_{u}||\Set{N}_{i}|}$, where $\Set{N}_{u}$ and $\Set{N}_{i}$ denote the first-hop neighbors of user $u$ and item $i$.
From the viewpoint of representation learning, $p_{ui}$ reflects how much the historical item contributes the user preference.
From the viewpoint of message passing, $p_{ui}$ can be interpreted as a discount factor, considering the messages being propagated should decay with the path length.


\vspace{+5pt}
\noindent{\textbf{Message Aggregation}}. In this stage, we aggregate the messages propagated from $u$'s neighborhood to refine $u$'s representation.
Specifically, we define the aggregation function as:
\begin{gather}\label{equ:1-aggregator}
	\Mat{e}^{(1)}_{u}=\text{LeakyReLU}\Big(\Mat{m}_{u \leftarrow u} + \sum_{i\in\Set{N}_{u}}\Mat{m}_{u\leftarrow i}\Big),
\end{gather}
where $\Mat{e}^{(1)}_{u}$ denotes the representation of user $u$ obtained after the first embedding propagation layer. 
The activation function of LeakyReLU~\cite{LeakyRelu} allows messages to encode both positive and small negative signals.
Note that in addition to the messages propagated from neighbors $\Set{N}_u$,  we take the self-connection of $u$ into consideration: $\Mat{m}_{u\leftarrow u}=\Mat{W}_1\Mat{e}_{u}$, which retains the information of original features ($\Mat{W}_1$ is the weight matrix shared with the one used in Equation (\ref{equ:0-message})).
Analogously, we can obtain the representation $\Mat{e}_{i}^{(1)}$ for item $i$ by propagating information from its connected users.
To summarize, the advantage of the embedding propagation layer lies in explicitly exploiting the first-order connectivity information to relate user and item representations.

\begin{figure}[t]
\centering
\includegraphics[width=0.42\textwidth]{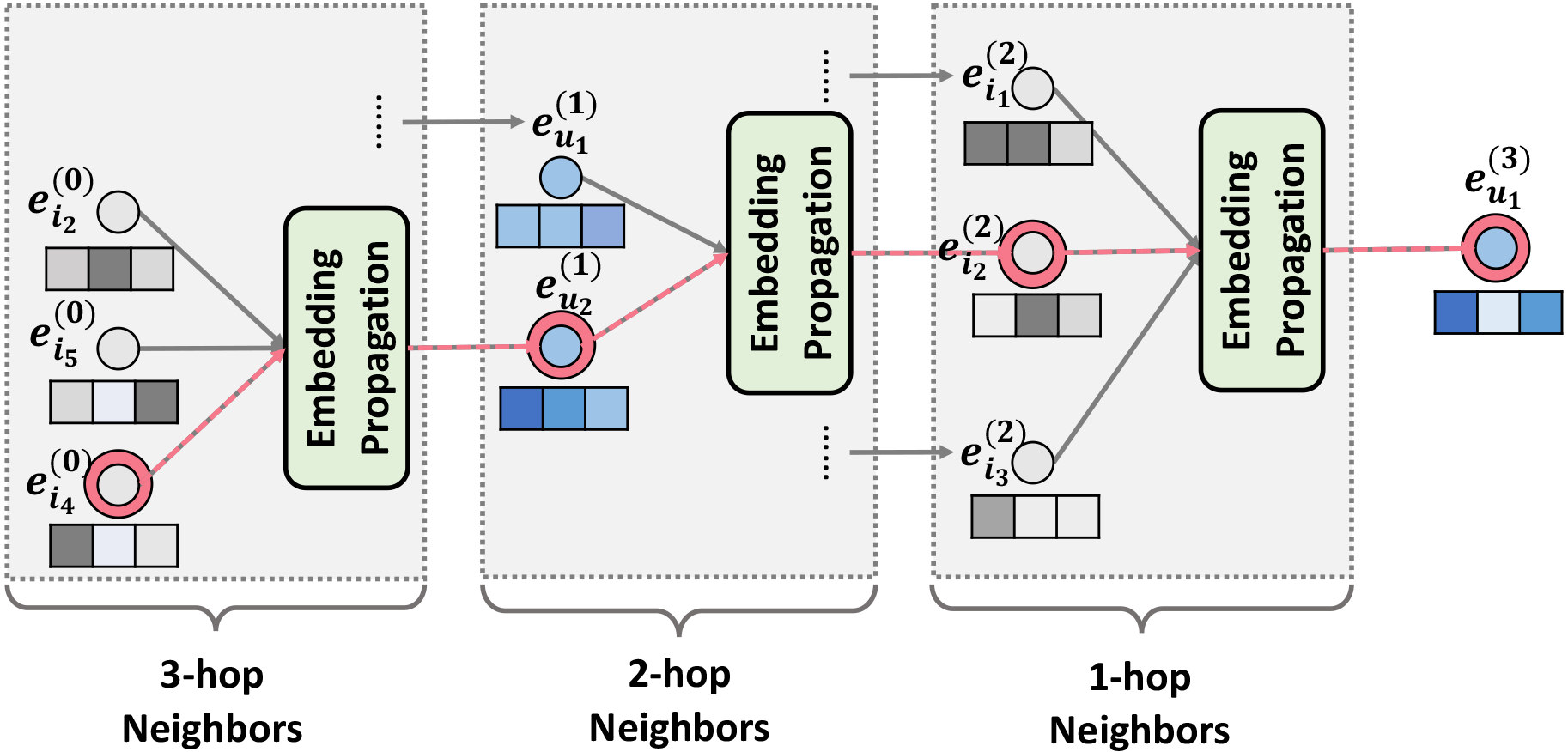}
\vspace{-10pt}
\caption{Illustration of third-order embedding propagation for user $u_{1}$. Best view in color. 
}
\label{fig:higher-order-propagation}
\vspace{-15pt}
\end{figure}

\vspace{5px}
\subsubsection{\textbf{High-order Propagation}}
With the representations aug-mented by first-order connectivity modeling, we can stack more embedding propagation layers to explore the high-order connectivity information. 
Such high-order connectivities are crucial to encode the collaborative signal to estimate the relevance score between a user and item.

By stacking $l$ embedding propagation layers, a user (and an item) is capable of receiving the messages propagated from its $l$-hop neighbors.
As Figure~\ref{fig:embedding-propagation} displays, in the $l$-th step, the representation of user $u$ is recursively formulated as:
\begin{gather}\label{equ:l-aggregator}
\Mat{e}_{u}^{(l)}=\text{LeakyReLU}\Big(\Mat{m}^{(l)}_{u\leftarrow u}+\sum_{i\in\Set{N}_{u}}\Mat{m}^{(l)}_{u\leftarrow i}\Big),
\end{gather}
wherein the messages being propagated are defined as follows,
\begin{gather}\label{equ:l-message}
\begin{cases}
\Mat{m}^{(l)}_{u\leftarrow i}=p_{ui}\Big(\Mat{W}^{(l)}_{1}\Mat{e}^{(l-1)}_{i} + \Mat{W}^{(l)}_{2}(\Mat{e}^{(l-1)}_{i}\odot\Mat{e}^{(l-1)}_{u})\Big),\\
\Mat{m}^{(l)}_{u\leftarrow u}=\Mat{W}^{(l)}_{1}\Mat{e}^{(l-1)}_{u},
\end{cases}
\end{gather}
where $\Mat{W}^{(l)}_{1},\Mat{W}^{(l)}_{2},\in\Space{R}^{d_{l}\times d_{l-1}}$ are the trainable transformation matrices, and $d_{l}$ is the transformation size;
$\Mat{e}_{i}^{(l-1)}$ is the item representation generated from the previous message-passing steps, memorizing the messages from its $(l$-1)-hop neighbors.
It further contributes to the representation of user $u$ at layer $l$.
Analogously, we can obtain the representation for item $i$ at the layer $l$.

As Figure~\ref{fig:higher-order-propagation} shows, the collaborative signal like $u_{1}\leftarrow i_{2}\leftarrow u_{2}\leftarrow i_{4}$ can be captured in the embedding propagation process.
Furthermore, the message from $i_{4}$ is explicitly encoded in $\Mat{e}^{(3)}_{u_{1}}$ (indicated by the red line).
As such, stacking multiple embedding propagation layers seamlessly injects collaborative signal into the representation learning process.

\vspace{5px}
\noindent\textbf{Propagation Rule in Matrix Form.}
To offer a holistic view of embedding propagation and facilitate batch implementation, we provide the matrix form of the layer-wise propagation rule (equivalent to Equations~\eqref{equ:l-aggregator} and~\eqref{equ:l-message}):
\begin{align}\label{equ:rule}
\Mat{E}^{(l)}=\text{LeakyReLU}\Big((\Lapl+\Mat{I})\Mat{E}^{(l-1)}\Mat{W}^{(l)}_{1} + \Lapl\Mat{E}^{(l-1)}\odot\Mat{E}^{(l-1)}\Mat{W}^{(l)}_{2}\Big),
\end{align}
where $\Mat{E}^{(l)}\in\Space{R}^{(N+M)\times d_{l}}$ are the representations for users and items obtained after $l$ steps of embedding propagation.
$\Mat{E}^{(0)}$ is set as $\Mat{E}$ at the initial message-passing iteration, that is $\Mat{e}^{(0)}_{u}=\Mat{e}_{u}$ and $\Mat{e}^{(0)}_{i}=\Mat{e}_{i}$;
and $\Mat{I}$ denote an identity matrix.
$\Lapl$ represents the Laplacian matrix for the user-item graph, which is formulated as:
\begin{align}
\Lapl=\Mat{D}^{-\frac{1}{2}}\Mat{A}\Mat{D}^{-\frac{1}{2}}~~\text{and}~~\Mat{A}=\BlockMatSquare{\Mat{0}}{\Mat{R}}{\Trans{\Mat{R}}}{\Mat{0}},
\end{align}
where $\Mat{R}\in\space{R}^{N\times M}$ is the user-item interaction matrix, and $\Mat{0}$ is all-zero matrix;
$\Mat{A}$ is the adjacency matrix
and $\Mat{D}$ is the diagonal degree matrix, where the $t$-th diagonal element $D_{tt}=|\Set{N}_{t}|$;
as such, the nonzero off-diagonal entry $\Lapl_{ui}=1/\sqrt{|\Set{N}_{u}||\Set{N}_{i}|}$, which is equal to $p_{ui}$ used in Equation~\eqref{equ:0-message}.

By implementing the matrix-form propagation rule, we can simultaneously update the representations for all users and items in a rather efficient way. It allows us to discard the node sampling procedure, which is commonly used to make graph convolution network runnable on large-scale graph~\cite{DeepInf}.
We will analyze the complexity in Section~\ref{sec:complexity}.

\subsection{Model Prediction}

After propagating with $L$ layers, we obtain multiple representations for user $u$, namely $\{\Mat{e}^{(1)}_{u},\cdots,\Mat{e}^{(L)}_{u}\}$.
Since the representations obtained in different layers emphasize the messages passed over different connections, they have different contributions in reflecting user preference.
As such, we concatenate them to constitute the final embedding for a user; we do the same operation on items, concatenating the item representations $\{\Mat{e}^{(1)}_{i},\cdots,\Mat{e}^{(L)}_{i}\}$ learned by different layers to get the final item embedding:
\begin{equation}\label{equ:final-rep}
\Mat{e}^{*}_{u}= \Mat{e}^{(0)}_{u}\Vert\cdots\Vert\Mat{e}^{(L)}_{u}, \quad
\Mat{e}^{*}_{i}=\Mat{e}^{(0)}_{i}\Vert\cdots\Vert\Mat{e}^{(L)}_{i},
\end{equation}
where $\Vert$ is the concatenation operation.
By doing so, we not only enrich the initial embeddings with embedding propagation layers, but also allow controlling the range of propagation by adjusting $L$. Note that besides concatenation, other aggregators can also be applied, such as weighted average, max pooling, LSTM, \etc, which imply different assumptions in combining the connectivities of different orders. The advantage of using concatenation lies in its simplicity, since it involves no additional parameters to learn, and it has been shown quite effectively in a recent work of graph neural networks~\cite{JumpKG}, which refers to layer-aggregation mechanism. 

Finally, we conduct the inner product to estimate the user's preference towards the target item:
\begin{align}
\hat{y}_{\text{NGCF}}(u,i)=\Trans{\Mat{e}^{*}_{u}}\Mat{e}^{*}_{i}.
\end{align}
In this work, we emphasize the embedding function learning thus only employ the simple interaction function of inner product. Other more complicated choices, such as neural network-based interaction functions~\cite{NCF}, are left to explore in the future work. 

\subsection{Optimization}
To learn model parameters, we optimize the pairwise BPR loss~\cite{BPRMF}, which has been intensively used in recommender systems~\cite{ACF,APR}. It considers the relative order between observed and unobserved user-item interactions.
Specifically, BPR assumes that the observed interactions, which are more reflective of a user's preferences, should be assigned higher prediction values than unobserved ones.
The objective function is as follows,
\begin{gather}\label{equ:loss}
	Loss = \sum_{(u,i,j)\in\Set{O}}-\ln\sigma(\hat{y}_{ui}-\hat{y}_{uj})+\lambda\norm{\Theta}^{2}_{2},
\end{gather}
where $\Set{O}=\{(u,i,j)|(u,i)\in\Set{R}^{+}, (u,j)\in\Set{R}^{-}\}$ denotes the pairwise training data, $\Set{R}^{+}$ indicates the observed interactions, and $\Set{R}^{-}$ is the unobserved interactions;
$\sigma(\cdot)$ is the sigmoid function;
$\Theta=\{\Mat{E}, \{\Mat{W}^{(l)}_{1}, \Mat{W}^{(l)}_{2}\}_{l=1}^L \}$ denotes all trainable model parameters, and $\lambda$ controls the $L_2$ regularization strength to prevent overfitting. We adopt mini-batch Adam~\cite{Adam} to optimize the prediction model and update the model parameters.
In particular, for a batch of randomly sampled triples $(u,i,j)\in\Set{O}$, we establish their representations $[\Mat{e}^{(0)},\cdots,\Mat{e}^{(L)}]$ after $L$ steps of propagation, and then update model parameters by using the gradients of the loss function.

\subsubsection{\textbf{Model Size.}} It is worth pointing out that although NGCF obtains an embedding matrix ($\textbf{E}^{(l)}$) at each propagation layer $l$, it only introduces very few parameters --- two weight matrices of size $d_l\times d_{l-1}$. Specifically, these embedding matrices are derived from the embedding look-up table $\textbf{E}^{(0)}$, with the transformation based on the user-item graph structure and weight matrices. 
As such, compared to MF --- the most concise embedding-based recommender model, our NGCF uses only $2L d_l d_{l-1}$ more parameters. 
Such additional cost on model parameters is almost negligible, considering that $L$ is usually a number smaller than 5, and $d_l$ is typically set as the embedding size, which is much smaller than the number of users and items. For example, on our experimented Gowalla dataset (20K users and 40K items), when the embedding size is 64 and we use $3$ propagation layers of size $64\times64$, MF has 4.5 million parameters, while our NGCF uses only 0.024 million additional parameters. To summarize, NGCF uses very few additional model parameters to achieve the high-order connectivity modeling. 


\subsubsection{\textbf{Message and Node Dropout.}}
Although deep learning models have strong representation ability, they usually suffer from overfitting.
Dropout is an effective solution to prevent neural networks from overfitting.
Following the prior work on graph convolutional network~\cite{GC-MC}, we propose to adopt two dropout techniques in NGCF: \emph{message dropout} and \emph{node dropout}.
Message dropout randomly drops out the outgoing messages.
Specifically, we drop out the messages being propagated in Equation~\eqref{equ:l-message}, with a probability $p_{1}$.
As such, in the $l$-th propagation layer, only partial messages contribute to the refined representations.
We also conduct node dropout to randomly block a particular node and discard all its outgoing messages.
For the $l$-th propagation layer, we randomly drop $(M+N)p_{2}$ nodes of the Laplacian matrix, where $p_{2}$ is the dropout ratio.

Note that dropout is only used in training, and must be disabled during testing. 
The message dropout endows the representations more robustness against the presence or absence of single connections between users and items, and the node dropout focuses on reducing the influences of particular users or items. 
We perform experiments to investigate the impact of message dropout and node dropout on NGCF in Section~\ref{sec:dropout}.

\subsection{Discussions}\label{ss:discussions}
In the subsection, we first show how NGCF generalizes SVD++~\cite{SVDFeatures}.
In what follows, we analyze the time complexity of NGCF.

\subsubsection{\textbf{NGCF Generalizes SVD++}}\label{sec:generalize-mf-svd}
SVD++ can be viewed as a special case of NGCF with no high-order propagation layer.
In particular, we set $L$ to one.
Within the propagation layer, we disable the transformation matrix and nonlinear activation function.
Thereafter, $\Mat{e}^{(1)}_{u}$ and $\Mat{e}^{(1)}_{i}$ are treated as the final representations for user $u$ and item $i$, respectively.
We term this simplified model as NGCF-SVD, which can be formulated as:
\begin{align}\label{equ:NGCF-svd}
\hat{y}_{\text{NGCF-SVD}}=\Trans{(\Mat{e}_{u}+\sum_{i'\in\Set{N}_{u}}p_{ui'}\Mat{e}_{i'})}(\Mat{e}_{i}+\sum_{u'\in\Set{N}_{i}}p_{iu'}\Mat{e}_{i}).
\end{align}
Clearly, by setting $p_{ui'}$ and $p_{u'i}$ as $1/\sqrt{|\Set{N}_{u}|}$ and $0$ separately, we can exactly recover SVD++ model.
Moreover, another widely-used item-based CF model, FISM~\cite{FISM}, can be also seen as a special case of NGCF, wherein $p_{iu'}$ in Equation~\eqref{equ:NGCF-svd} is set as $0$.

\subsubsection{\textbf{Time Complexity Analysis}}\label{sec:complexity}
As we can see, the layer-wise propagation rule is the main operation.
For the $l$-th propagation layer, the matrix multiplication has computational complexity $O(|\Set{R}^{+}|d_{l}d_{l-1})$, where $|\Set{R}^{+}|$ denotes the number of nonzero entires in the Laplacian matrix; and $d_{l}$ and $d_{l-1}$ are the current and previous transformation size.
For the prediction layer, only the inner product is involved, for which the time complexity of the whole training epoch is $O(\sum_{l=1}^{L}|\Set{R}^{+}|d_{l})$.
Therefore, the overall complexity for evaluating NGCF is $O(\sum_{l=1}^{L}|\Set{R}^{+}|d_{l}d_{l-1}+\sum_{l=1}^{L}|\Set{R}^{+}|d_{l})$.
Empirically, under the same experimental settings (as explained in Section~\ref{sec:experiments}), MF and NGCF cost around $20s$ and $80s$ per epoch on Gowalla dataset for training, respectively;
during inference, the time costs of MF and NGCF are $80s$ and $260s$ for all testing instances, respectively.
\section{Related Work}\label{sec:related-work}
We review existing work on model-based CF, graph-based CF, and graph neural network-based methods, which are most relevant with this work.
Here we highlight the differences with our NGCF.


\subsection{Model-based CF Methods}
Modern recommender systems~\cite{NCF,CMN,TEM} parameterize users and items by vectorized representations and reconstruct user-item interaction data based on model parameters.
For example, MF~\cite{MF,BPRMF} projects the ID of each user and item as an embedding vector, and conducts inner product between them to predict an interaction.
To enhance the embedding function, much effort has been devoted to incorporate side information like item content~\cite{CDL,ACF}, social relations~\cite{ItemSilk}, item relations~\cite{RCF}, user reviews~\cite{DBLP:conf/www/ChengDZK18}, and external knowledge graph~\cite{KGAT,KPRN}.
While inner product can force user and item embeddings of an observed interaction close to each other, its linearity makes it insufficient to reveal the complex and nonlinear relationships between users and items~\cite{NCF,CML}.
Towards this end, recent efforts~\cite{AutoCF,NCF,CML,NFM} focus on exploiting deep learning techniques to enhance the interaction function, so as to capture the nonlinear feature interactions between users and items.
For instance, neural CF models, such as NeuMF~\cite{NCF}, employ nonlinear neural networks as the interaction function;
meanwhile, translation-based CF models, such as LRML~\cite{tay2018latent}, instead model the interaction strength with Euclidean distance metrics.

Despite great success, we argue that the design of the embedding function is insufficient to yield optimal embeddings for CF, since the CF signals are only implicitly captured.
Summarizing these methods, the embedding function transforms the descriptive features (\eg ID and attributes) to vectors, while the interaction function serves as a similarity measure on the vectors.
Ideally, when user-item interactions are perfectly reconstructed, the transitivity property of behavior similarity could be captured.
However, such transitivity effect showed in the Running Example is not explicitly encoded, thus there is no guarantee that the indirectly connected users and items are close in the embedding space.
Without an explicit encoding of the CF signals, it is hard to obtain embeddings that meet the desired properties. 

\subsection{Graph-Based CF Methods}
Another line of research~\cite{RecWalk,BiRank,HOP-rec} exploits the user-item interaction graph to infer user preference.
Early efforts, such as ItemRank~\cite{ItemRank} and BiRank~\cite{BiRank}, adopt the idea of label propagation to capture the CF effect.
To score items for a user, these methods define the labels as her interacted items, and propagate the labels on the graph. 
As the recommendation scores are obtained based on the structural reachness (which can be seen as a kind of similarity) between the historical items and the target item, these methods essentially belong to neighbor-based methods.
However, these methods are conceptually inferior to model-based CF methods, since there lacks model parameters to optimize the objective function of recommendation. 

The recently proposed method HOP-Rec~\cite{HOP-rec} alleviates the problem by combining graph-based with embedding-based method.
It first performs random walks to enrich the interactions of a user with multi-hop connected items.
Then it trains MF with BPR objective based on the enriched user-item interaction data to build the recommender model. 
The superior performance of HOP-Rec over MF provides evidence that incorporating the connectivity information is beneficial to obtain better embeddings in capturing the CF effect.
However, we argue that HOP-Rec does not fully explore the high-order connectivity, which is only utilized to enrich the training data\footnote{The enriched trained data can be seen as a regularizer to the original training.}, rather than directly contributing to the model's embedding function. Moreover, the performance of HOP-Rec depends heavily on the random walks, which require careful tuning efforts such as a proper setting of decay factor.  


\subsection{Graph Convolutional Networks}
By devising a specialized graph convolution operation on user-item interaction graph~(\cf Equation~(\ref{equ:0-message})), we make NGCF effective in exploiting the CF signal in high-order connectivities. Here we discuss existing recommendation methods that also employ graph convolution operations~\cite{GC-MC,PinSage,SpectralCF}. 


GC-MC~\cite{GC-MC} applies the graph convolution network (GCN)~\cite{GCN} on user-item graph, however it only employs one convolutional layer to exploit the direct connections between users and items.
Hence it fails to reveal collaborative signal in high-order connectivities.
PinSage~\cite{PinSage} is an industrial solution that employs multiple graph convolution layers on item-item graph for Pinterest image recommendation. As such, the CF effect is captured on the level of item relations, rather than the collective user behaviors.
SpectralCF~\cite{SpectralCF} proposes a spectral convolution operation to discover all possible connectivity between users and items in the spectral domain.
Through the eigen-decomposition of graph adjacency matrix, it can discover the connections between a user-item pair.
However, the eigen-decomposition causes a high computational complexity, which is very time-consuming and difficult to support large-scale recommendation scenarios.

\section{Experiments}\label{sec:experiments}
We perform experiments on three real-world datasets to evaluate our proposed method, especially the embedding propagation layer.
We aim to answer the following research questions:
\begin{itemize}[leftmargin=*]
\item \textbf{RQ1}: How does NGCF perform as compared with state-of-the-art CF methods?
\item \textbf{RQ2}: How do different hyper-parameter settings (\eg depth of layer, embedding propagation layer, layer-aggregation mechanism, message dropout, and node dropout) affect NGCF?
\item \textbf{RQ3}: How do the representations benefit from the high-order connectivity?
\end{itemize}

\begin{table}[t]
\caption{Statistics of the datasets.}
\vspace{-10px}
\label{tab:dataset}
\resizebox{0.42\textwidth}{!}{
\begin{tabular}{l|r|r|r|r}
\hline
\multicolumn{1}{c|}{Dataset} & \multicolumn{1}{c|}{\#Users} & \multicolumn{1}{c|}{\#Items} & \multicolumn{1}{c|}{\#Interactions} & \multicolumn{1}{c}{Density} \\ \hline\hline
Gowalla & $29,858$ & $40,981$ & $1,027,370$ & $0.00084$ \\ \hline
Yelp2018$^{*}$ & $31,668$ & $38,048$ & $1,561,406$ & $0.00130$ \\ \hline
Amazon-Book & $52,643$ & $91,599$ & $2,984,108$ & $0.00062$ \\ \hline
\end{tabular}}
\vspace{-15px}
\end{table}

\subsection{Dataset Description}
To evaluate the effectiveness of NGCF, we conduct experiments on three benchmark datasets: Gowalla, Yelp2018$^{*}$\footnote{In the previous version of yelp2018, we did not filter out cold-start items in the testing set, and hence we rerun all methods.}, and Amazon-book, which are publicly accessible and vary in terms of domain, size, and sparsity.
We summarize the statistics of three datasets in Table~\ref{tab:dataset}.

\vspace{2px}
\noindent\textbf{Gowalla:}
This is the check-in dataset~\cite{gowalla} obtained from Gowalla, where users share their locations by checking-in.
To ensure the quality of the dataset, we use the $10$-core setting~\cite{VBPR}, \ie retaining users and items with at least ten interactions.

\vspace{2px}
\noindent\textbf{Yelp2018$^{*}$:}
This dataset is adopted from the 2018 edition of the Yelp challenge.
Wherein, the local businesses like restaurants and bars are viewed as the items.
We use the same $10$-core setting in order to ensure data quality.

\vspace{2px}
\noindent\textbf{Amazon-book:}
Amazon-review is a widely used dataset for product recommendation~\cite{amazon-review}.
We select Amazon-book from the collection.
Similarly, we use the $10$-core setting to ensure that each user and item have at least ten interactions.

For each dataset, we randomly select $80\%$ of historical interactions of each user to constitute the training set, and treat the remaining as the test set.
From the training set, we randomly select $10\%$ of interactions as validation set to tune hyper-parameters.
For each observed user-item interaction, we treat it as a positive instance, and then conduct the negative sampling strategy to pair it with one negative item that the user did not consume before.

\subsection{Experimental Settings}
\subsubsection{\textbf{Evaluation Metrics}}
For each user in the test set, we treat all the items that the user has not interacted with as the negative items.
Then each method outputs the user's preference scores over all the items, except the positive ones used in the training set.
To evaluate the effectiveness of top-$K$ recommendation and preference ranking, we adopt two widely-used evaluation protocols~\cite{NCF,HOP-rec}: recall@$K$ and ndcg@$K$\footnote{The previous implementation of ndcg metric in NGCF follows the codes~\cite{IRGAN}, but we find it is slightly different from the standard definition, although reflecting the similar trends. We re-implement the ndcg metric and rerun all methods}.
By default, we set $K=20$.
We report the average metrics for all users in the test set.



\subsubsection{\textbf{Baselines}}
To demonstrate the effectiveness, we compare our proposed NGCF with the following methods:

\begin{itemize}[leftmargin=*]
\item \textbf{MF}~\cite{BPRMF}: This is matrix factorization optimized by the Bayesian personalized ranking (BPR) loss, which exploits the user-item direct interactions only as the target value of interaction function.

\item \textbf{NeuMF}~\cite{NCF}: The method is a state-of-the-art neural CF model which uses multiple hidden layers above the element-wise and concatenation of user and item embeddings to capture their non-linear feature interactions.
Especially, we employ two-layered plain architecture, where the dimension of each hidden layer keeps the same.


\item \textbf{CMN}~\cite{CMN}: It is a state-of-the-art memory-based model, where the user representation attentively combines the memory slots of neighboring users via the memory layers.
Note that the first-order connections are used to find similar users who interacted with the same items.

\item \textbf{HOP-Rec}~\cite{HOP-rec}: This is a state-of-the-art graph-based model, where the high-order neighbors derived from random walks are exploited to enrich the user-item interaction data.

\item \textbf{PinSage}~\cite{PinSage}: PinSage is designed to employ GraphSAGE~\cite{GraphSAGE} on item-item graph. In this work, we apply it on user-item interaction graph. Especially, we employ two graph convolution layers as suggested in~\cite{PinSage}, and the hidden dimension is set equal to the embedding size.

\item \textbf{GC-MC}~\cite{GC-MC}: This model adopts GCN~\cite{GCN} encoder to generate the representations for users and items, where only the first-order neighbors are considered. Hence one graph convolution layer, where the hidden dimension is set as the embedding size, is used as suggested in~\cite{GC-MC}.
\end{itemize}
We also tried SpectralCF~\cite{SpectralCF} but found that the eigen-decomposition leads to high time cost and resource cost, especially when the number of users and items is large.
Hence, although it achieved promising performance in small datasets, we did not select it for comparison.
For fair comparison, all methods optimize the BPR loss as shown in Equation~\eqref{equ:loss}.

\begin{figure*}[t]
\centering
\subfigure[ndcg on Gowalla]{
\label{fig:sparsity-gowalla}\includegraphics[width=0.31\textwidth]{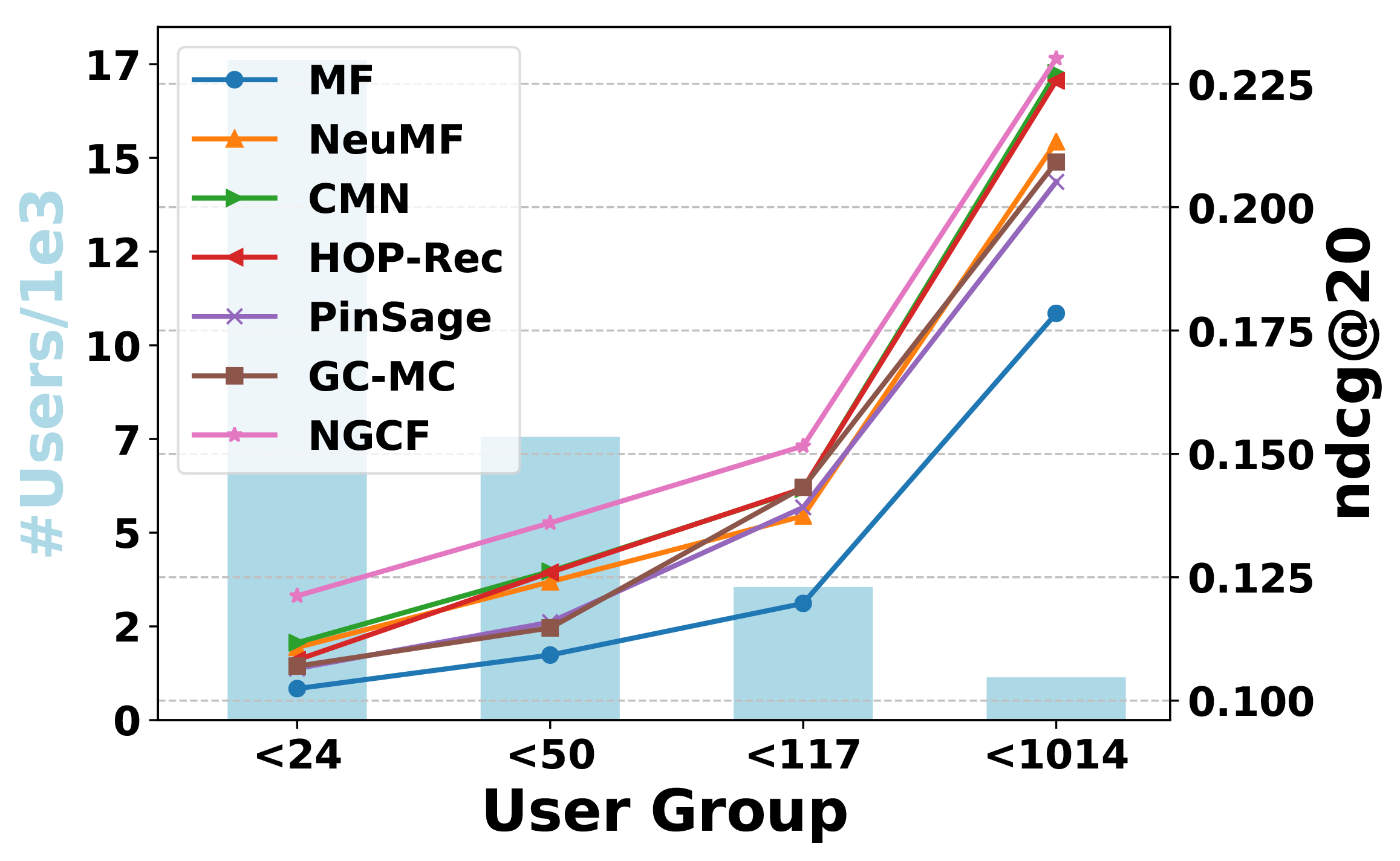}}
\subfigure[ndcg on Yelp2018$^{*}$]{
\label{fig:sparsity-yelp2018}\includegraphics[width=0.31\textwidth]{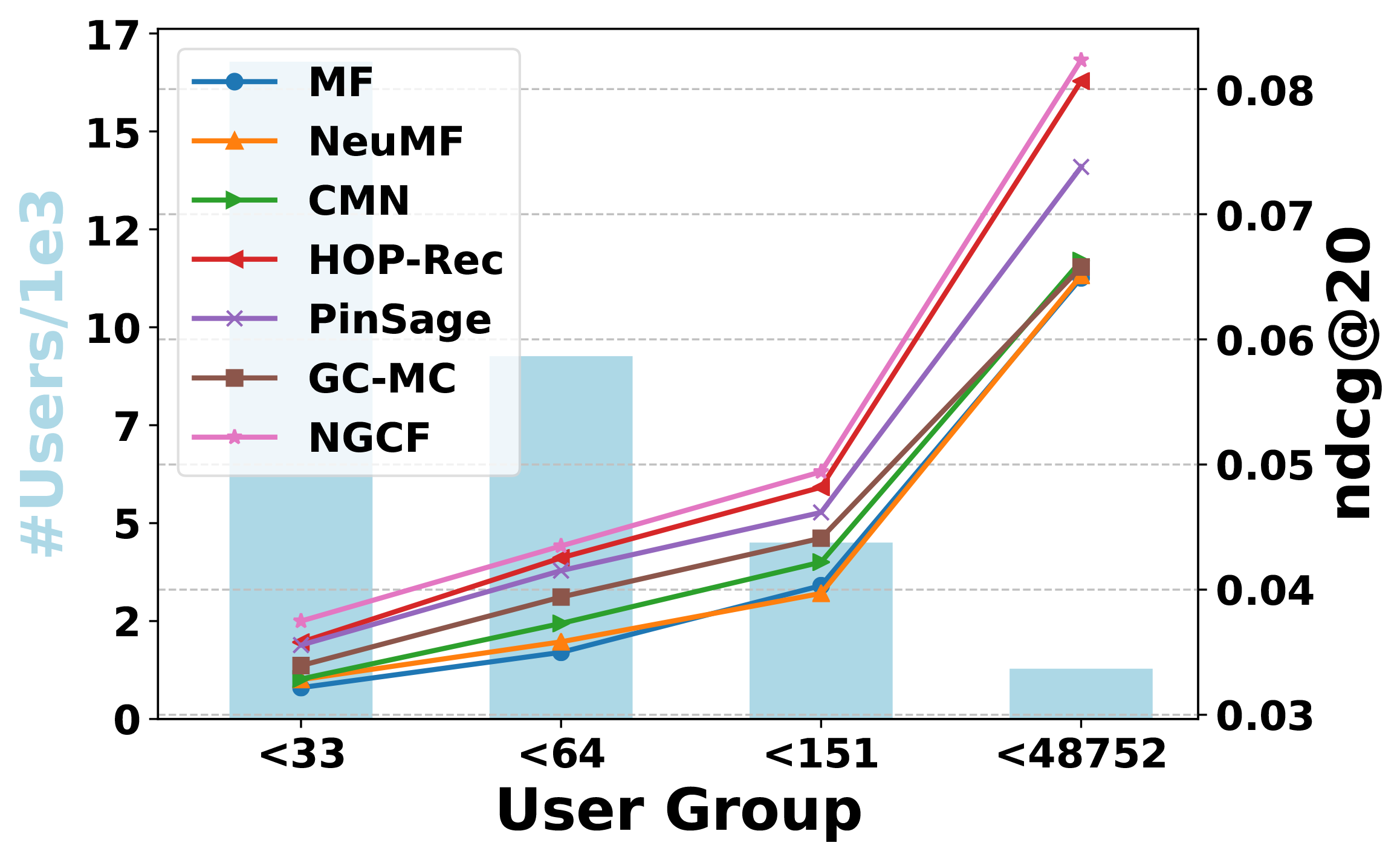}}
\subfigure[ndcg on Amazon-book]{
\label{fig:sparsity-amazon-book}\includegraphics[width=0.31\textwidth]{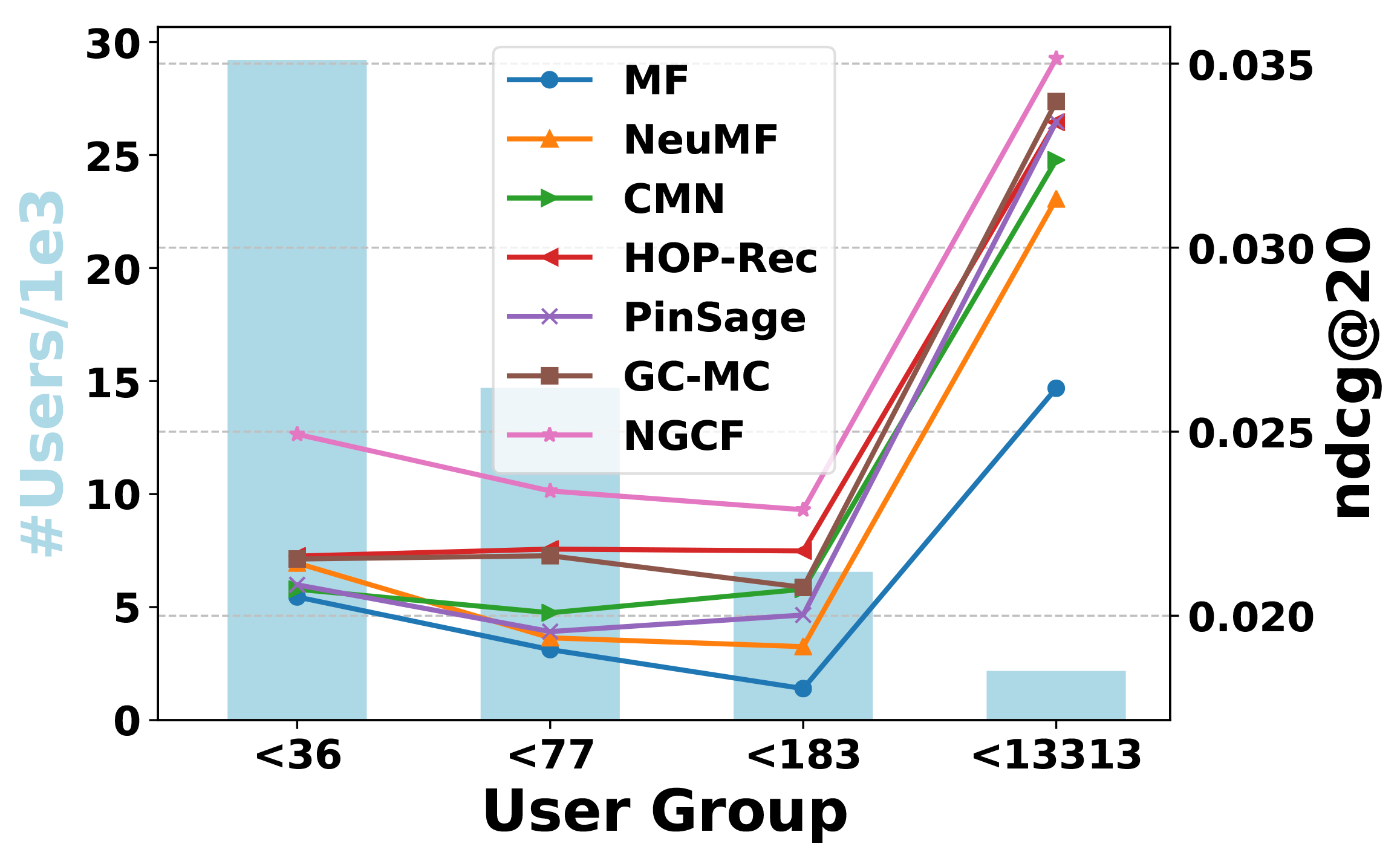}}
\vspace{-15pt}
\caption{Performance comparison over the sparsity distribution of user groups on different datasets.
Wherein, the background histograms indicate the number of users involved in each group, and the lines demonstrate the performance \wrt ndcg@$20$.}
\label{fig:sparsity}\vspace{-10pt}
\end{figure*}

\subsubsection{\textbf{Parameter Settings}}
We implement our NGCF model in Tensorflow.
The embedding size is fixed to $64$ for all models.
For HOP-Rec, we search the steps of random walks in $\{1,2,3\}$ and tune the learning rate in $\{0.025,0.020,0.015,0.010\}$.
We optimize all models except HOP-Rec with the Adam optimizer, where the batch size is fixed at $1024$.
In terms of hyperparameters, we apply a grid search for hyperparameters: the learning rate is tuned amongst $\{0.0001,0.0005,0.001,0.005\}$, the coefficient of $L_{2}$ normalization is searched in $\{10^{-5},10^{-4},\cdots,10^{1},10^{2}\}$, and the dropout ratio in $\{0.0,0.1,\cdots,0.8\}$.
Besides, we employ the node dropout technique for GC-MC and NGCF, where the ratio is tuned in $\{0.0,0.1,\cdots,0.8\}$.
We use the Xavier initializer~\cite{Xarvier} to initialize the model parameters\footnote{We train MF from scratch, and use the trained MF embeddings to initialize NeuMF, GC-MC, PinSage, and NGCF to speed up and stabilize the training process. Since we update the implementation of BPR loss, we rerun the experiments.}.
Moreover, early stopping strategy is performed, \ie premature stopping if recall@$20$ on the validation data does not increase for $50$ successive epochs.
To model the CF signal encoded in third-order connectivity, we set the depth of NGCF $L$ as three.
Without specification, we show the results of three embedding propagation layers, node dropout ratio of $0.0$, and message dropout ratio of $0.1$.

\subsection{Performance Comparison (RQ1)}
We start by comparing the performance of all the methods, and then explore how the modeling of high-order connectivity improves under the sparse settings.

\begin{table}[t]
\caption{Overall Performance Comparison.}
\vspace{-10px}
\label{tab:overall-performance}
\resizebox{0.48\textwidth}{!}{
\begin{tabular}{l|c c|c c|c c}
\hline
 & \multicolumn{2}{c|}{Gowalla} & \multicolumn{2}{c|}{Yelp2018$^{*}$} & \multicolumn{2}{c}{Amazon-Book} \\ 
 & recall & ndcg & recall & ndcg & recall & ndcg \\ \hline\hline
MF      & 0.1291 & 0.1109 & 0.0433 & 0.0354 & 0.0250 & 0.0196 \\ 
NeuMF   & 0.1399 & 0.1212 & 0.0451 & 0.0363 & 0.0258 & 0.0200 \\
CMN     & $\underline{0.1405}$ & $\underline{0.1221}$ & 0.0457 & 0.0369 & 0.0267 & 0.0218 \\ \hline
HOP-Rec & 0.1399 & 0.1214 & $\underline{0.0517}$ & $\underline{0.0428}$ & $\underline{0.0309}$ & $\underline{0.0232}$ \\ \hline
GC-MC   & 0.1395 & 0.1204 & 0.0462 & 0.0379 & 0.0288 & 0.0224  \\
PinSage & 0.1380 & 0.1196 & 0.0471 & 0.0393 & 0.0282 & 0.0219  \\ \hline
\textbf{NGCF-3} & $\Mat{0.1569^{*}}$ & $\Mat{0.1327^{*}}$ & $\Mat{0.0579^{*}}$ & $\Mat{0.0477^{*}}$ & $\Mat{0.0337^{*}}$ & $\Mat{0.0261^{*}}$ \\ \hline\hline
\%Improv. & 11.68\% & 8.64\% & 11.97\% & 11.29\% & 9.61\% & 12.50\% \\ 
$p$-value & 2.01e-7 & 3.03e-3 & 5.34e-3 & 4.62e-4 & 3.48e-5 & 1.26e-4 \\\hline
\end{tabular}}
\vspace{-15px}
\end{table}

\subsubsection{\textbf{Overall Comparison}}
Table~\ref{tab:overall-performance} reports the performance comparison results.
We have the following observations:
\begin{itemize}[leftmargin=*]
\item MF achieves poor performance on three datasets. This indicates that the inner product is insufficient to capture the complex relations between users and items, further limiting the performance.
NeuMF consistently outperforms MF across all cases, demonstrating the importance of nonlinear feature interactions between user and item embeddings.
However, neither MF nor NeuMF explicitly models the connectivity in the embedding learning process, which could easily lead to suboptimal representations.

\item Compared to MF and NeuMF, the performance of GC-MC verifies that incorporating the first-order neighbors can improve the representation learning.
However, in Yelp2018$^{*}$, GC-MC underperforms NeuMF \wrt ndcg@$20$.
The reason might be that GC-MC fails to fully explore the nonlinear feature interactions between users and items.


\item CMN generally achieves better performance than GC-MC in most cases.
Such improvement might be attributed to the neural attention mechanism, which can specify the attentive weight of each neighboring user, rather than the equal or heuristic weight used in GC-MC.

\item PinSage slightly underperforms CMN in Gowalla and Amazon-Book, while performing much better in Yelp2018$^{*}$;
meanwhile, HOP-Rec generally achieves remarkable improvements in most cases.
It makes sense since PinSage introduces high-order connectivity in the embedding function, and HOP-Rec exploits high-order neighbors to enrich the training data, while CMN considers the similar users only.
It therefore points to the positive effect of modeling the high-order connectivity or neighbors.

\item NGCF consistently yields the best performance on all the datasets.
In particular, NGCF improves over the strongest baselines \wrt recall@$20$ by $11.68\%$, $11.97\%$, and $9.61\%$ in Gowalla, Yelp2018$^{*}$, and Amazon-Book, respectively.
By stacking multiple embedding propagation layers, NGCF is capable of exploring the high-order connectivity in an explicit way, while CMN and GC-MC only utilize the first-order neighbors to guide the representation learning.
This verifies the importance of capturing collaborative signal in the embedding function.
Moreover, compared with PinSage, NGCF considers multi-grained representations to infer user preference, while PinSage only uses the output of the last layer.
This demonstrates that different propagation layers encode different information in the representations.
And the improvements over HOP-Rec indicate that explicit encoding CF in the embedding function can achieve better representations.
We conduct one-sample t-tests and $p$-value $<0.05$ indicates that the improvements of NGCF over the strongest baseline (underlined) are statistically significant.

\end{itemize}

\subsubsection{\textbf{Performance Comparison \wrt Interaction Sparsity Levels}}\label{sec:sparsity}


The sparsity issue usually limits the expressiveness of recommender systems, since few interactions of inactive users are insufficient to generate high-quality representations.
We investigate whether exploiting connectivity information helps to alleviate this issue.

Towards this end, we perform experiments over user groups of different sparsity levels.
In particular, based on interaction number per user, we divide the test set into four groups, each of which has the same total interactions.
Taking Gowalla dataset as an example, the interaction numbers per user are less than $24$, $50$, $117$, and $1014$ respectively.
Figure~\ref{fig:sparsity} illustrates the results \wrt ndcg@$20$ on different user groups in Gowalla, Yelp2018$^{*}$, and Amazon-Book; 
we see a similar trend for performance \wrt recall@$20$ and omit the part due to the space limitation.
We find that:
\begin{itemize}[leftmargin=*]
\item NGCF and HOP-Rec consistently outperform all other baselines on all user groups.
It demonstrates that exploiting high-order connectivity greatly facilitates the representation learning for inactive users, as the collaborative signal can be effectively captured.
Hence, it might be promising to solve the sparsity issue in recommender systems, and we leave it in future work.

\item Jointly analyzing Figures~\ref{fig:sparsity-gowalla},~\ref{fig:sparsity-yelp2018}, and ~\ref{fig:sparsity-amazon-book}, we observe that the improvements achieved in the first two groups (\eg $8.49\%$ and $7.79\%$ over the best baselines separately for $<24$ and $<50$ in Gowalla) are more significant than that of the others (\eg $1.29\%$ for $<1014$ Gowalla groups).
It verifies that the embedding propagation is beneficial to the relatively inactive users.
\end{itemize}

\subsection{Study of NGCF (RQ2)}
As the embedding propagation layer plays a pivotal role in NGCF, we investigate its impact on the performance.
We start by exploring the influence of layer numbers.
We then study how the Laplacian matrix (\ie discounting factor $p_{ui}$ between user $u$ and item $i$) affects the performance.
Moreover, we analyze the influences of key factors, such as node dropout and message dropout ratios.
We also study the training process of NGCF.

\subsubsection{\textbf{Effect of Layer Numbers}}


\begin{table}[t]
\caption{Effect of embedding propagation layer numbers ($L$).
}
\vspace{-10px}
\label{tab:depth}
\resizebox{0.44\textwidth}{!}{
\begin{tabular}{l|c c|c c|c c}
\hline
 & \multicolumn{2}{c|}{Gowalla} & \multicolumn{2}{c|}{Yelp2018$^{*}$} & \multicolumn{2}{c}{Amazon-Book} \\ 
 & recall & ndcg & recall & ndcg & recall & ndcg \\ \hline\hline
NGCF-1 & 0.1556 & 0.1315 & 0.0543 & 0.0442 & 0.0313 & 0.0241 \\ 
NGCF-2 & 0.1547 & 0.1307 & 0.0566 & 0.0465 & 0.0330 & 0.0254\\ 
NGCF-3 & 0.1569 & 0.1327 & \textbf{0.0579} & \textbf{0.0477} & 0.0337 & 0.0261 \\ 
NGCF-4 & \textbf{0.1570} & \textbf{0.1327} & 0.0566 & 0.0461 & \textbf{0.0344} & \textbf{0.0263} \\ \hline
\end{tabular}}
\vspace{-10px}
\end{table}


To investigate whether NGCF can benefit from multiple embedding propagation layers, we vary the model depth.
In particular, we search the layer numbers in the range of $\{1,2,3,4\}$.
Table~\ref{tab:depth} summarizes the experimental results, wherein NGCF-3 indicates the model with three embedding propagation layers, and similar notations for others.
Jointly analyzing Tables~\ref{tab:overall-performance} and~\ref{tab:depth}, we have the following observations:
\begin{itemize}[leftmargin=*]
\item Increasing the depth of NGCF substantially enhances the recommendation cases.
Clearly, NGCF-2 and NGCF-3 achieve consistent improvement over NGCF-1 across all the board, which considers the first-order neighbors only.
We attribute the improvement to the effective modeling of CF effect: collaborative user similarity and collaborative signal are carried by the second- and third-order connectivities, respectively.

\item When further stacking propagation layer on the top of NGCF-3, we find that NGCF-4 leads to overfitting on Yelp2018$^{*}$ dataset.
This might be caused by applying a too deep architecture might introduce noises to the representation learning.
The marginal improvements on the other two datasets verifies that conducting three propagation layers is sufficient to capture the CF signal.

\item When varying the number of propagation layers, NGCF is consistently superior to other methods across three datasets.
It again verifies the effectiveness of NGCF, empirically showing that explicit modeling of high-order connectivity can greatly facilitate the recommendation task.

\end{itemize}

\subsubsection{\textbf{Effect of Embedding Propagation Layer and Layer-Aggregation Mechanism}}\label{sec:layer-effect}

To investigate how the embedding propagation (\ie graph convolution) layer affects the performance, we consider the variants of NGCF-1 that use different layers.
In particular, we remove the representation interaction between a node and its neighbor from the message passing function (\cf Equation~\eqref{equ:0-message}) and set it as that of PinSage and GC-MC, termed NGCF-1$_{\text{PinSage}}$ and NGCF-1$_{\text{GC-MC}}$ respectively.
Moreover, following SVD++, we obtain one variant based on Equations~\eqref{equ:NGCF-svd}, termed NGCF-1$_{\text{SVD++}}$.
We show the results in Table~\ref{tab:laplacian} and have the following findings:

\begin{itemize}[leftmargin=*]
    \item NGCF-1 is consistently superior to all variants. We attribute the improvements to the representation interactions (\ie $\Mat{e}_{u}\odot\Mat{e}_{i}$), which makes messages being propagated dependent on the affinity between $\Mat{e}_i$ and $\Mat{e}_u$ and functions like the attention mechanism~\cite{ACF}. Whereas, all variants only take linear transformation into consideration. It hence verifies the rationality and effectiveness of our embedding propagation function.
    
    \item In most cases, NGCF-1$_{\text{SVD++}}$ underperforms NGCF-1$_{\text{PinSage}}$ and NGCF-1$_{\text{GC-MC}}$. It illustrates the importance of messages passed by the nodes themselves and the nonlinear transformation.
    
    \item Jointly analyzing Tables~\ref{tab:overall-performance} and~\ref{tab:laplacian}, we find that, when concatenating all layers' outputs together, NGCF-1$_{\text{PinSage}}$ and NGCF-1$_{\text{GC-MC}}$ achieve better performance than PinSage and GC-MC, respectively.
    This emphasizes the significance of layer-aggregation mechanism, which is consistent with~\cite{JumpKG}.
\end{itemize}

\begin{table}[t]
\caption{Effect of graph convolution layers.
}
\vspace{-10px}
\label{tab:laplacian}
\resizebox{0.45\textwidth}{!}{
\begin{tabular}{l|c c|c c|c c}
\hline
 & \multicolumn{2}{c|}{Gowalla} & \multicolumn{2}{c|}{Yelp2018$^{*}$} & \multicolumn{2}{c}{Amazon-Book} \\ 
 & recall & ndcg & recall & ndcg & recall & ndcg \\ \hline\hline
NGCF-1 & $\Mat{0.1556}$ & $\Mat{0.1315}$ & $\Mat{0.0543}$ & $\Mat{0.0442}$ & $\Mat{0.0313}$ & $\Mat{0.0241}$ \\ 
NGCF-1$_{\text{SVD++}}$ & 0.1517 & 0.1265 & 0.0504 & 0.0414 & 0.0297 & 0.0232\\ 
NGCF-1$_{\text{GC-MC}}$ & 0.1523 & 0.1307 & 0.0518 & 0.0420 & 0.0305 & 0.0234 \\ 
NGCF-1$_{\text{PinSage}}$   & 0.1534 & 0.1308 & 0.0516 & 0.0420 & 0.0293 & 0.0231 \\ \hline
\end{tabular}}
\vspace{-10px}
\end{table}

\subsubsection{\textbf{Effect of Dropout}}\label{sec:dropout}
\begin{figure}[t]
\centering
\subfigure[Gowalla]{
\label{fig:dropout-gowalla}\includegraphics[width=0.151\textwidth]{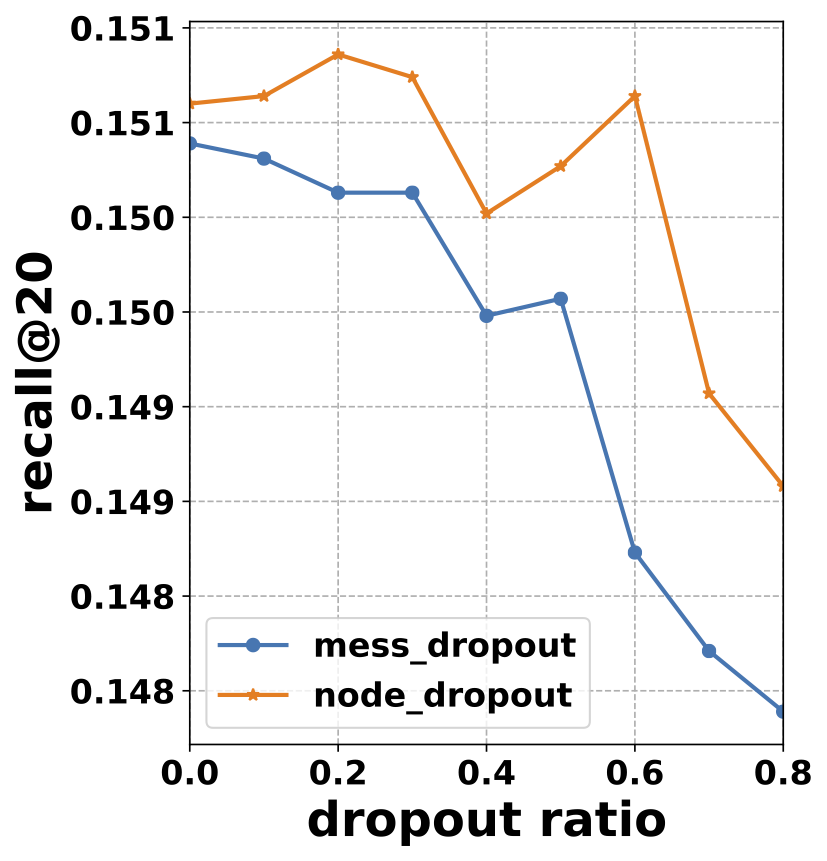}}
\subfigure[Yelp2018$^{*}$]{
\label{fig:dropout-yelp2018}\includegraphics[width=0.151\textwidth]{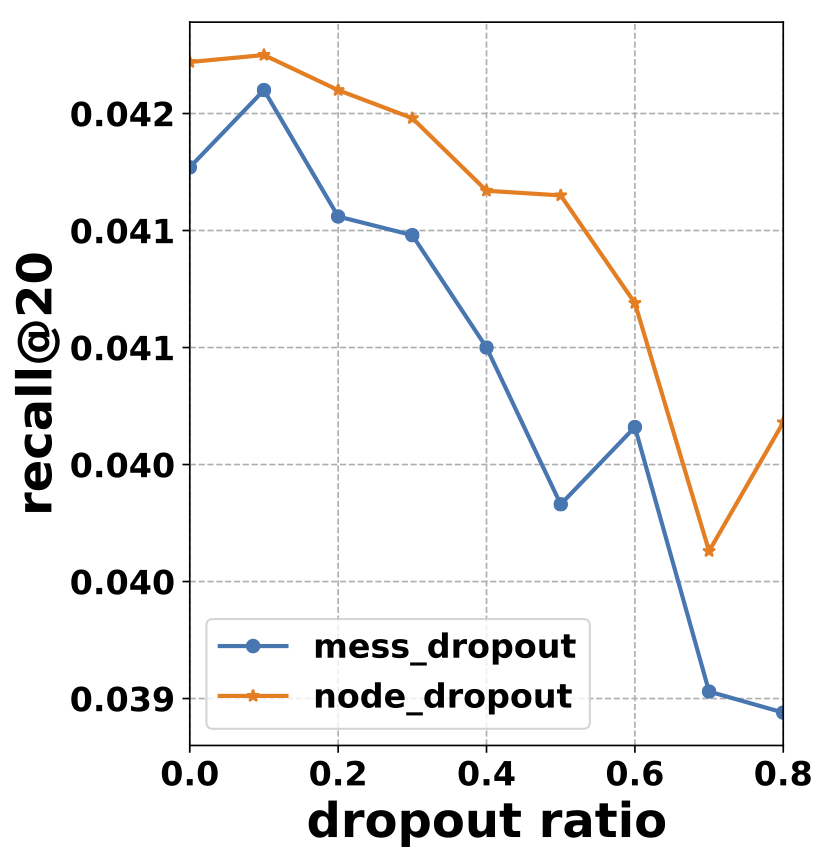}}
\subfigure[Amazon-Book]{
\label{fig:gamma-ndcg-mi}\includegraphics[width=0.151\textwidth]{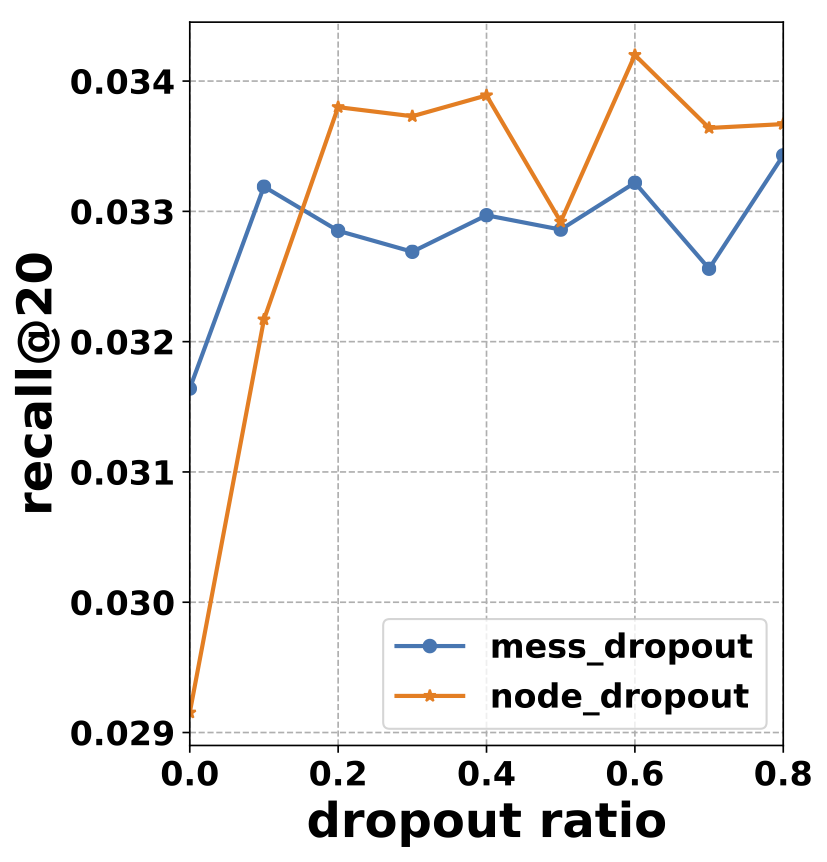}}
\vspace{-10pt}
\caption{Effect of node dropout and message dropout ratios.
}
\label{fig:dropout}\vspace{-15pt}
\end{figure}

Following the prior work~\cite{GC-MC}, we employ node dropout and message dropout techniques to prevent NGCF from overfitting.
Figure~\ref{fig:dropout} plots the effect of message dropout ratio $p_{1}$ and node dropout ratio $p_{2}$ against different evaluation protocols on different datasets.

Between the two dropout strategies, node dropout offers better performance.
Taking Gowalla as an example, setting $p_{2}$ as $0.2$ leads to the highest recall@$20$ of $0.1514$, which is better than that of message dropout $0.1506$.
One reason might be that dropping out all outgoing messages from particular users and items makes the representations robust against not only the influence of particular edges, but also the effect of nodes.
Hence, node dropout is more effective than message dropout, which is consistent with the findings of prior effort~\cite{GC-MC}.
We believe this is an interesting finding, which means that node dropout can be an effective strategy to address overfitting of graph neural networks.

\subsubsection{\textbf{Test Performance \wrt Epoch}}

\begin{figure}[t]
\centering
\subfigure[Gowalla]{
\label{fig:trend-gowalla}\includegraphics[width=0.15\textwidth]{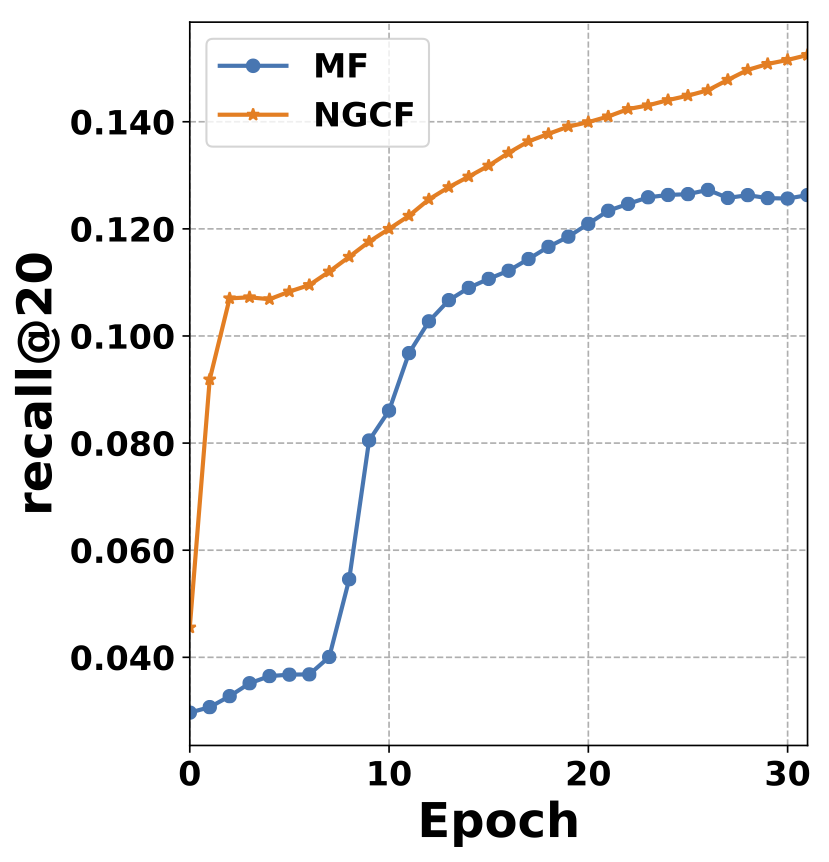}}
\subfigure[Yelp2018$^{*}$]{
\label{fig:trend-yelp2018}\includegraphics[width=0.15\textwidth]{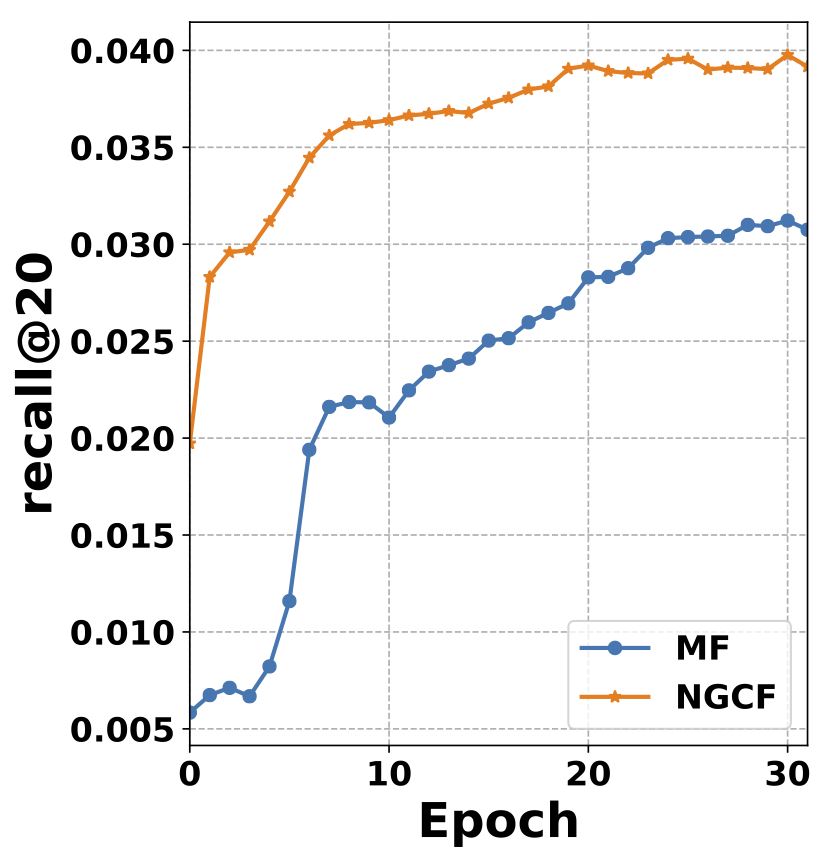}}
\subfigure[Amazon-Book]{
\label{fig:trend-amazon-book}\includegraphics[width=0.15\textwidth]{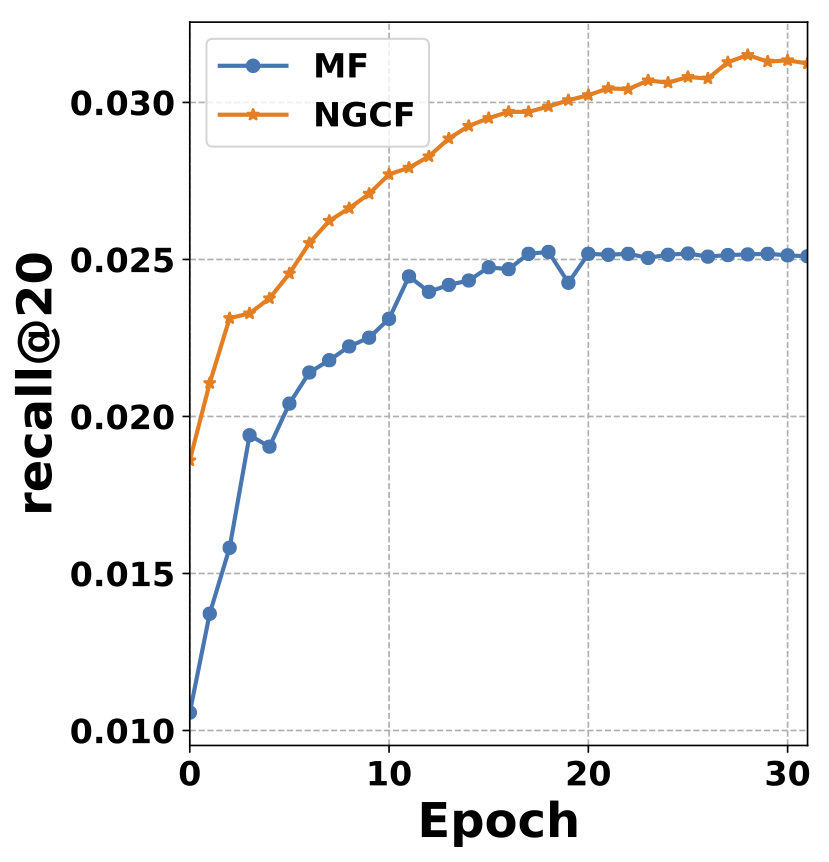}}
\vspace{-10pt}
\caption{Test performance of each epoch of MF and NGCF.}
\label{fig:trend}\vspace{-15pt}
\end{figure}

Figure~\ref{fig:trend} shows the test performance \wrt recall of each epoch of MF and NGCF. Due to the space limitation, we omit the performance \wrt ndcg which has the similar trend.
We can see that, NGCF exhibits fast convergence than MF on three datasets.
It is reasonable since indirectly connected users and items are involved when optimizing the interaction pairs in mini-batch.
Such an observation demonstrates the better model capacity of NGCF and the effectiveness of performing embedding propagation in the embedding space.

\subsection{Effect of High-order Connectivity (RQ3)}

In this section, we attempt to understand how the embedding propagation layer facilitates the representation learning in the embedding space.
Towards this end, we randomly selected six users from Gowalla dataset, as well as their relevant items.
We observe how their representations are influenced \wrt the depth of NGCF.

\begin{figure}[t]
\centering
\subfigure[MF (NGCF-0)]{
\label{fig:rep-bprmf}\includegraphics[width=0.23\textwidth]{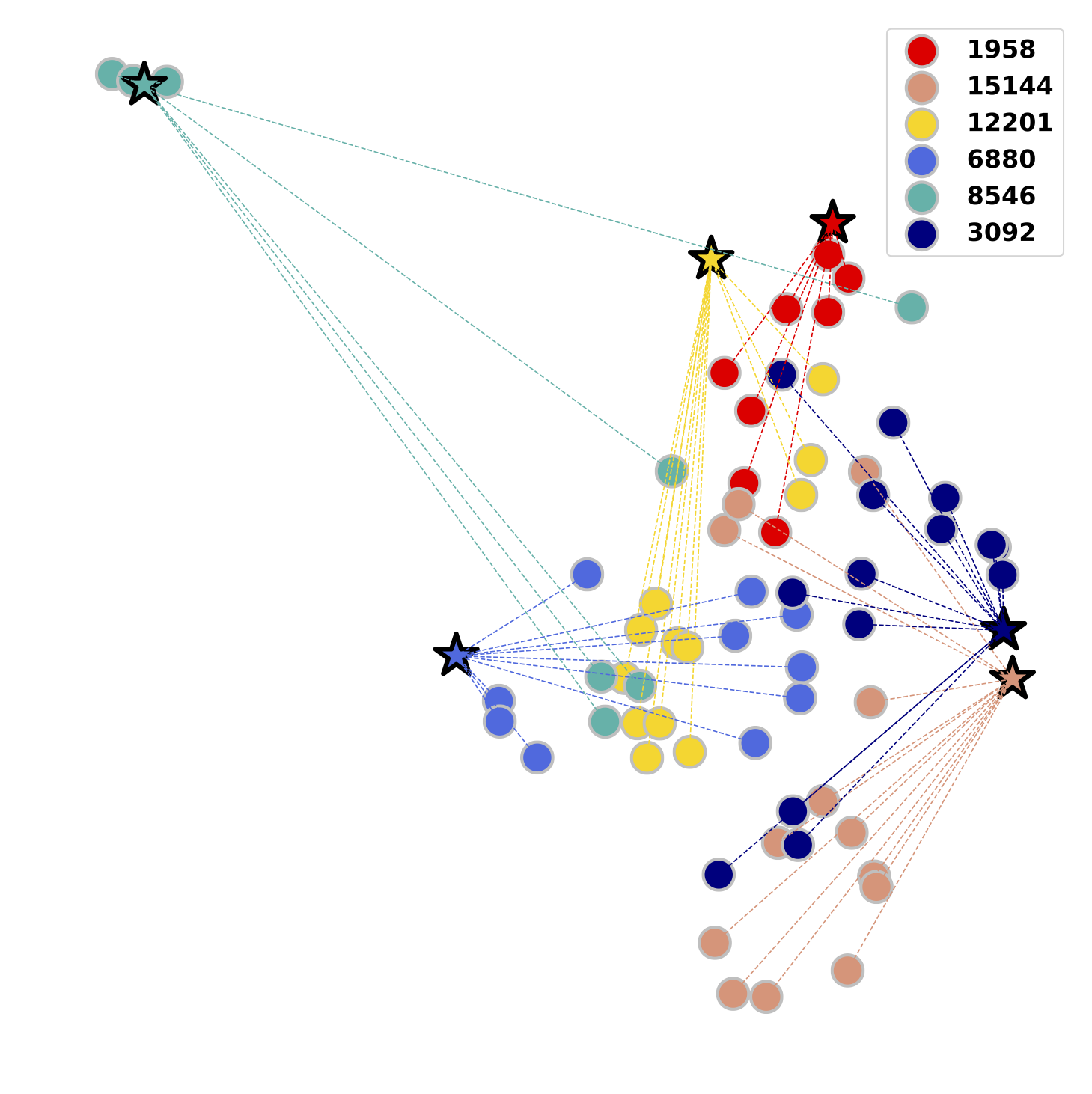}}
\subfigure[NGCF-3]{
\label{fig:rep-depr3}\includegraphics[width=0.23\textwidth]{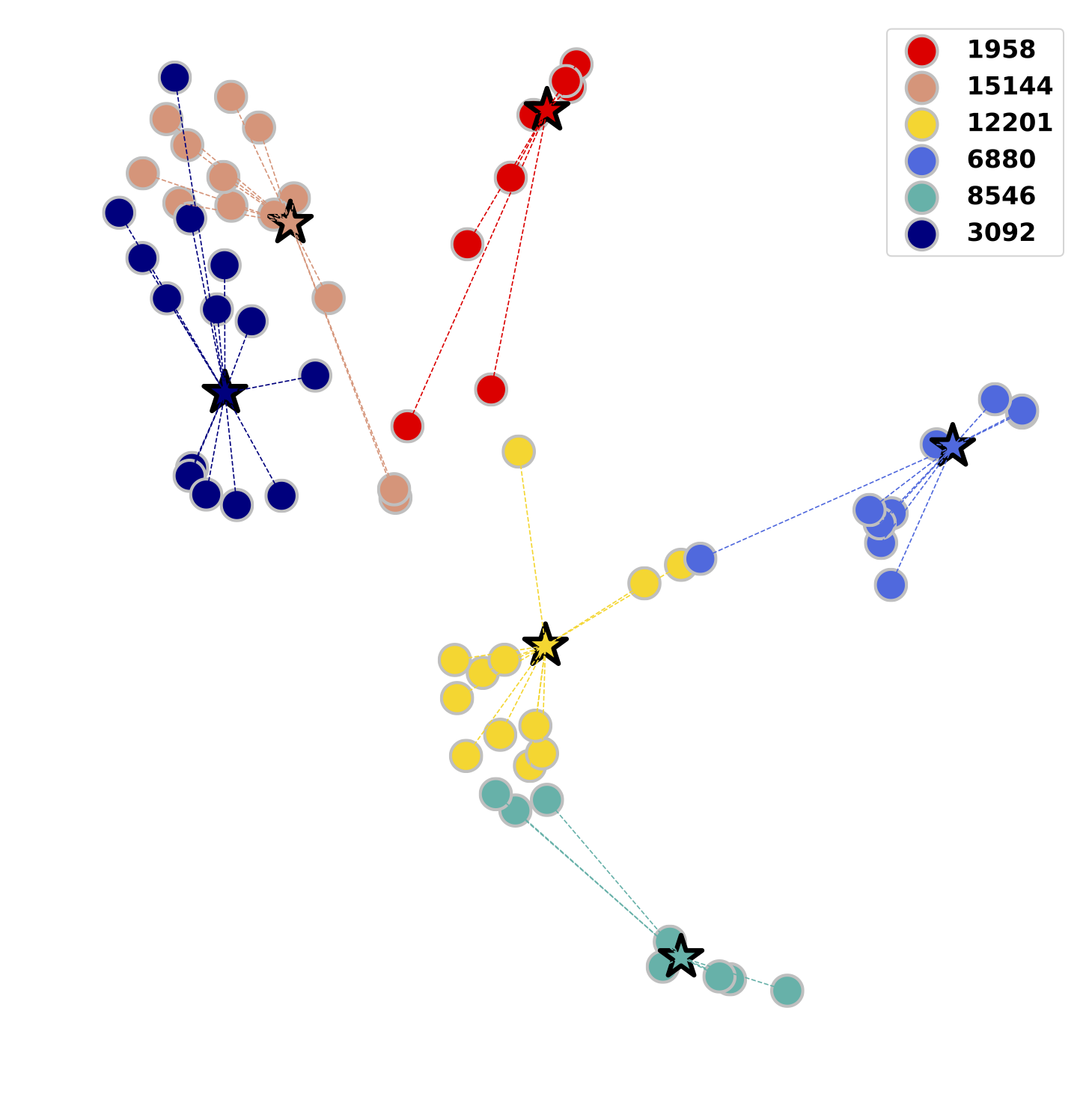}}
\vspace{-10pt}
\caption{Visualization of the learned t-SNE transformed representations derived from MF and NGCF-3.
Each star represents a user from Gowalla dataset, while the points with the same color denote the relevant items. Best view in color.}
\label{fig:gamma-effect}\vspace{-15pt}
\end{figure}

Figures~\ref{fig:rep-bprmf} and~\ref{fig:rep-depr3} show the visualization of the representations derived from MF (\ie NGCF-0) and NGCF-3, respectively.
Note that the items are from the test set, which are not paired with users in the training phase.
There are two key observations:
\begin{itemize}[leftmargin=*]
\item The connectivities of users and items are well reflected in the embedding space, that is, they are embedded into the near part of the space.
In particular, the representations of NGCF-3 exhibit discernible clustering, meaning that the points with the same colors (\ie the items consumed by the same users) tend to form the clusters.

\item Jointly analyzing the same users (\eg $12201$ and $6880$) across Figures~\ref{fig:rep-bprmf} and~\ref{fig:rep-depr3}, we find that, when stacking three embedding propagation layers, the embeddings of their historical items tend to be closer.
It qualitatively verifies that the proposed embedding propagation layer is capable of injecting the explicit collaborative signal (via NGCF-3) into the representations.

\end{itemize}

\section{Conclusion and Future Work}

In this work, we explicitly incorporated collaborative signal into the embedding function of model-based CF.
We devised a new framework NGCF, which achieves the target by leveraging high-order connectivities in the user-item integration graph.
The key of NGCF is the newly proposed embedding propagation layer, based on which we allow the embeddings of users and items interact with each other to harvest the collaborative signal.
Extensive experiments on three real-world datasets demonstrate the rationality and effectiveness of injecting the user-item graph structure into the embedding learning process. In future, we will further improve NGCF by incorporating the attention mechanism~\cite{ACF} to learn variable weights for neighbors during embedding propagation and for the connectivities of different orders. This will be beneficial to model generalization and interpretability. 
Moreover, we are interested in exploring the adversarial learning~\cite{APR} on user/item embedding and the graph structure for enhancing the robustness of NGCF.

This work represents an initial attempt to exploit structural knowledge with the message-passing mechanism in model-based CF and opens up new research possibilities.
Specifically, there are many other forms of structural information can be useful for understanding user behaviors, such as the cross features~\cite{FashionInterpretable} in context-aware and semantics-rich recommendation~\cite{DBLP:conf/mm/LiuWZSXYL17,SongFHYLN18}, item knowledge graph~\cite{KGAT}, and social networks~\cite{ItemSilk}.
For example, by integrating item knowledge graph with user-item graph, we can establish knowledge-aware connectivities between users and items, which help unveil user decision-making process in choosing items.
We hope the development of NGCF is beneficial to the reasoning of user online behavior towards more effective and interpretable recommendation.

\vspace{5px}
\noindent\textbf{Acknowledgement:}
This research is part of NExT++ research and also supported by the Thousand Youth Talents Program 2018.
NExT++ is supported by the National Research Foundation, Prime Minister's Office, Singapore under its IRC@SG Funding Initiative.



\bibliographystyle{ACM-Reference-Format}
\balance

\bibliography{ms}
\balance

\end{document}